# Optical ReLU-like Activation Function Based on a Semiconductor Laser with Optical Injection

GUAN-TING LIU,[1] YI-WEI SHEN,[1] RUI-QIAN LI,[1] JINGYI YU,[1] XUMING HE,[1] AND CHENG WANG[1,2,*]

[1]School of Information Science and Technology, ShanghaiTech University, 201210 Shanghai, China
[2]Shanghai Engineering Research Center of Energy Efficient and Custom AI IC, ShanghaiTech University, Shanghai 201210, China
*Corresponding author: wangcheng1@shanghaitech.edu.cn

**Artificial neural networks usually consist of successive linear multiply-accumulate operations and nonlinear activation functions. However, most optical neural networks only achieve the linear operation in the optical domain, while the optical implementation of activation function remains challenging. Here we present an optical ReLU-like activation function based on a semiconductor laser subject to the optical injection in experiment. The ReLU-like function is achieved in a broad regime above the Hopf bifurcation of the injection-locking diagram. In particular, the slope of the activation function is reconfigurable by tuning the frequency difference between the master laser and the slave laser.**

The explosive growth of artificial neural networks (ANNs) has been substantially advancing the development of artificial intelligence. Meanwhile, the ANNs demand a huge amount of computing power, which is challenging the digital hardware based on the von Neumann architecture. Optical implementation of the ANNs is a promising solution to raise the computing power, which is known as optical neural networks (ONNs). ONNs are featured with inherent high bandwidth, high multiplexing dimensions, high energy efficiency as well as low latency [1-5]. In recent years, a large variety of ONNs have been successfully demonstrated, relying on Mach-Zehnder interferometers (MZI) [6], microring resonators (MRR) [7, 8], phase-changing materials [9, 10], diffraction units [11, 12], VCSEL arrays [13], etc. However, most ONNs only implement the linear multiply-accumulate operations (synapses) in the optical domain, while the nonlinear activation functions (neuron responses) are mostly implemented in the digital domain. This suggests both the optical-electrical conversion and the analog-digital conversion are required between each hidden layer of the ONNs. It is obvious that both conversions introduce large power consumption and high latency. In order to remove the analog-digital conversion, optoelectronic neurons have been demonstrated, which usually consist of a photodetector and an optical modulator [14, 15].

In comparison with the optoelectronic neurons, all-optical activation functions are more desirable, since no optical-electrical conversion is required. Shi *et al.* employed a semiconductor optical amplifier (SOA)-based wavelength converter to achieve a third-order polynomial nonlinear function, where a tunable laser was required to assist the wavelength conversion [16]. In addition, Mougrias-Alexandris *et al.* reported a sigmoid activation function using a SOA-based MZI, where three lasers were involved to achieve the nonlinear function [6]. Li *et al.* demonstrated a ReLU function based on a thin-film lithium niobate waveguide, where the second harmonic generation effect and the optical parametric amplification effect were exploited [17]. Jha *et al.* proposed a reconfigurable nonlinear activation function based on a MRR-loaded MZI, where sigmoid, ReLU, radial basis, and softplus functions were demonstrated [18]. In addition, Wu *et al.* proved radial basis, ReLU, and ELU functions based on a MRR, which used the thermal-optic effect [19]. In particular, both the excitability and the bi-stability of semiconductor lasers with optical injection were theoretically proposed to produce nonlinear activation functions, respectively [20, 21]. In experiment, a membrane laser with optical injection was shown to exhibit a ReLU function, when the laser was biased slightly below the lasing threshold [22]. On the other hand, a semiconductor laser produced a tanh-like function, when the optical injection is operated in the vicinity of the saddle-node bifurcation [23]. In this work, we present an all-optical ReLU-like function based on a semiconductor laser subject to the optical injection in experiment. The ReLU-like function is achieved in a broad regime above the Hopf bifurcation of the injection-locking diagram. In particular, the slope of the ReLU-like function is adjustable through tuning the frequency difference between the master laser and the slave laser. In addition, we prove that the performance of the optical ReLU-like function is comparable to that of the standard ReLU function.

Figure 1(a) shows the experimental setup for the operation of the all-optical ReLU-like activation function. The slave laser is a single-mode distributed feedback laser with an emission wavelength of $\lambda_s$, which acts as the nonlinear neuron. The master laser is a tunable external-cavity laser (Santec TSL-710) with a wavelength of $\lambda_m$. The light of the master laser is uni-directionally injected into the slave laser through an optical circulator. The polarization of the master laser is aligned with that of the slave laser through the polarization controller. The input power $P_{in}$ of the light injected into the slave laser is tracked by power meter 1. The output light of the slave laser passes through a bandpass filter (8 GHz bandwidth), and the center wavelength is set to be identical as the master laser ($\lambda_m$). Therefore, both the input signal and the output signal share the same wavelength. The output power $P_{out}$ of the filtered light is monitored by power meter 2. In the experiment, the slave laser is operated at 20 ℃, and the lasing threshold is $I_{th}$ = 8 mA. When biased at $1.9 \times I_{th}$, the output power of the free-running slave laser is 1.9 mW, and the emission wavelength is $\lambda_s$=1549.5 nm. It is known that the nonlinear dynamics of the slave laser with optical injection are primarily determined by the injection ratio and the detuning frequency [24]. The injection ratio is defined as the power ratio of the master laser to the slave laser, and the detuning frequency ($\Delta f$) is defined as the lasing frequency difference between both lasers. Figure 1(b) shows the measured injection-locking diagram formed by the input power $P_{in}$ (or injection ratio) and the detuning frequency. The stable locking regime is bounded by the Hopf bifurcation at the positive detuning side and the saddle-node (SN) bifurcation at the negative detuning side [25]. It is shown that the stable locking regime broadens with increasing injection strength. The laser produces continuous wave inside the stable regime. However, the laser produces rich nonlinear dynamics both above the Hopf bifurcation and below the SN bifurcation, including period-one oscillations [26], chaotic oscillations [27], and quasi-periodic oscillations [24].

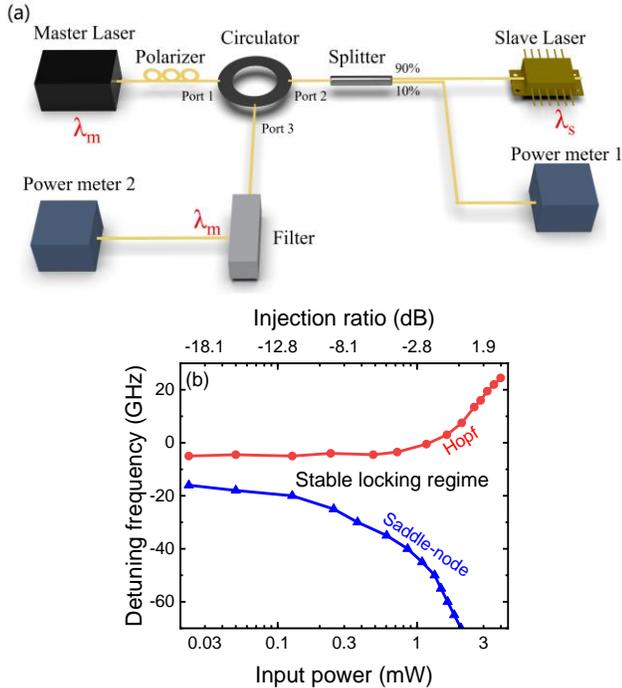

Fig. 1. (a) Experimental setup for the optical activation function. (b) Injection-locking diagram of the slave laser.

Within the stable locking regime, the lasing wavelength of the slave laser is locked to be the same as the master laser. Meanwhile, the phase of the slave laser is synchronized with the master laser as well [24]. Consequently, the optical spectrum of the injection-locked laser remains single mode (solid line), as shown in Fig. 2(a). Figure 2(b) shows that the output power $P_{out}$ of the slave laser responds almost linearly with the input power $P_{in}$ of the master laser, for different detuning frequencies. Consequently, this stable regime is not suitable for the operation of nonlinear activation function. Below the SN bifurcation at the negative detuning side, the slave laser becomes unlocked with the master laser. Thus, the optical spectrum at the detuning frequency of $\Delta f$=-30 GHz becomes dual modes in Fig. 2(c). The beating between the two modes results in sinusoidal wave-like period-one oscillations. This dual-mode emission in the optical domain and the periodic oscillation in the temporal domain may limit its application as a nonlinear neuron. In addition, in order to achieve a deep ONN with cascading optical neurons, both the input signal and the output signal of every neuron must maintain the same wavelength [28]. Therefore, it is crucial to apply a filter to pass only the optical wave of the master laser as shown in Fig. 1(a). For the detuning frequency of $\Delta f$=-30 GHz in Fig. 2(d), the output power increases almost linearly with the input power below the SN bifurcation (left side of the dash line), and the slope of the response curve is steep. Once the input power crosses the SN bifurcation (dash line), the slave laser is locked by the master laser and the injection-locked laser becomes a single mode as in Fig. 2(a). Meanwhile, the output power abruptly transits to a high level. Above the SN bifurcation (right side of the dash line), the output power responds almost linearly with the input power, which is the same as in Fig. 2(b). However, the curve slope becomes much smaller than the one below the SN bifurcation. Interestingly, the output power at the SN bifurcation either transits to a higher level or a lower level, depending on the detuning frequency. This is because the optical injection reduces the gain and raises the refractive index of the gain material [25]. Therefore, the cavity resonance shifts to the longer wavelength. When enlarging the detuning frequency from -20 GHz to -60 GHz, the lasing mode of the injection-locked laser in the stable regime becomes far away from the cavity resonance. As a result, less energy is extracted from the gain material and the output power becomes lower. In particular, the response curve at $\Delta f$=-40 GHz is similar to a leaky ReLU function but with 180° rotation. Nevertheless, the response curves for other detuning frequencies are discontinuous, which limits the application as nonlinear activation functions.

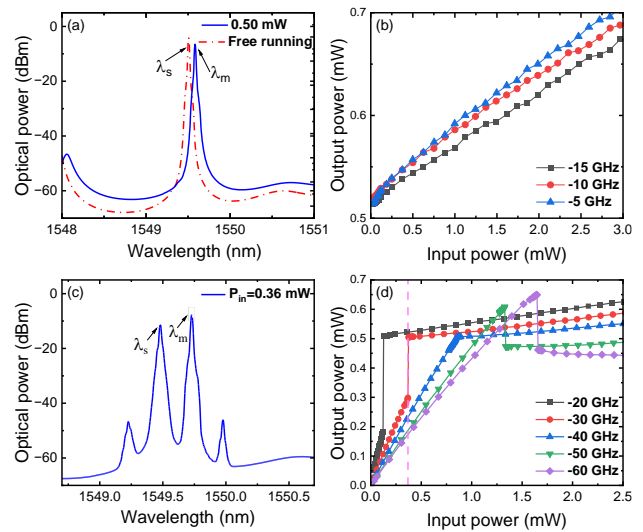

Fig. 2. (a) Optical spectrum within the stable regime. $\Delta f$=-10 GHz. (b) Optical power response within the stable regime. (c) Optical spectra below the saddle-node bifurcation. $\Delta f$=-30 GHz. (d) Optical power response in the vicinity of the SN bifurcation. The dash line indicates the position of the SN bifurcation at $\Delta f$=-30 GHz.

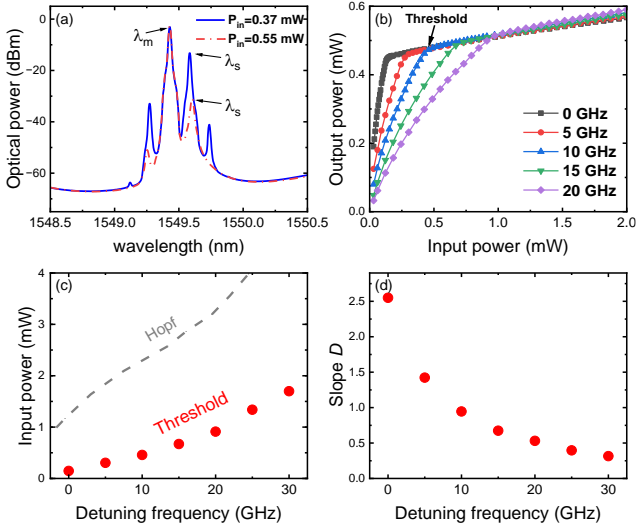

Fig. 3. (a) Optical spectra above the Hopf bifurcation. The solid line is below the threshold and the dash line is above the threshold. $\Delta f$=10 GHz. (b) ReLU-like activation functions at various detuning frequencies above the Hopf bifurcation. (c) Threshold power and (d) slope of the activation function versus the detuning frequency.

Above the Hopf bifurcation at the positive detuning side, Fig. 3(a) shows that the slave laser with optical injection shows dual lasing modes at the detuning frequency of $\Delta f$=10 GHz. The power difference between the master laser mode and the slave laser mode enlarges when increasing the input power from 0.37 mW (solid line) to 0.55 mW (dash line). Interestingly, the power response curve at $\Delta f$=10 GHz in Fig. 3(b) exhibits a kink at the input power of 0.46 mW. The shape of the response curve is like a leaky ReLU function with 180° rotation (or ReLU-like function for simplicity hereafter). We define the input power at the kink as the threshold power of the activation function. Below the threshold power, the power difference between the two modes is less than 15 dB (10.2 dB at 0.37 mW in Fig. 3(a)), leading to a strong power competition between both modes. In addition, the phases of both modes are almost un-correlated. When raising the input power, the master laser mode consumes more population inversion while the slave laser mode consumes less. Therefore, the output power of the master laser mode increases rapidly with increasing input power. Meanwhile, the amplitude of the slave laser mode declines. Above the threshold power, the power difference between the two modes is more than 15 dB (21.4 dB at 0.55 mW in Fig. 3(a)), that is, the slave laser mode is well suppressed. Moreover, the phases of both modes are partially correlated, which leads to strong frequency pushing effect [29, 30]. Then, almost all the carrier population contributes only to the master laser mode. As a result, the output power rises slowly with increasing input power. Interestingly, similar power evolution trend has been observed numerically, in the analysis of the period-one oscillation behavior [31]. The physical mechanism is attributed to the energy exchange between the master laser mode and the slave laser mode [31, 32]. In addition to the detuning frequency of 10 GHz, Fig. 3(b) demonstrates that the ReLU-like function occurs at various detuning frequencies. That is, the nonlinear activation function exists in a broad regime above the Hopf bifurcation, which is highly flexible for practical implementation. Figure 3(c) shows that the threshold power rises nonlinearly from 0.15 mW to 1.7 mW, when increasing the detuning frequency from 0 GHz to 30 GHz. In particular, the slope of the ReLU-like function below threshold is reconfigurable through adjusting the detuning frequency. The slope $D$ in Fig. 3(d) decreases nonlinearly from 2.55 at $\Delta f$=0 GHz down to 0.32 at $\Delta f$=30 GHz. On the other hand, the slope of the function above threshold in Fig. 3(b) is almost constant around 0.065. We remark that our experiment proves that the slave laser biased at other currents produces ReLU-like function as well, as long as it is above the lasing threshold. Below the lasing threshold, nevertheless, the power response of the slave laser is similar to the saturation behavior of a semiconductor optical amplifier [33], which is unlike the response of the membrane laser reported in [22].

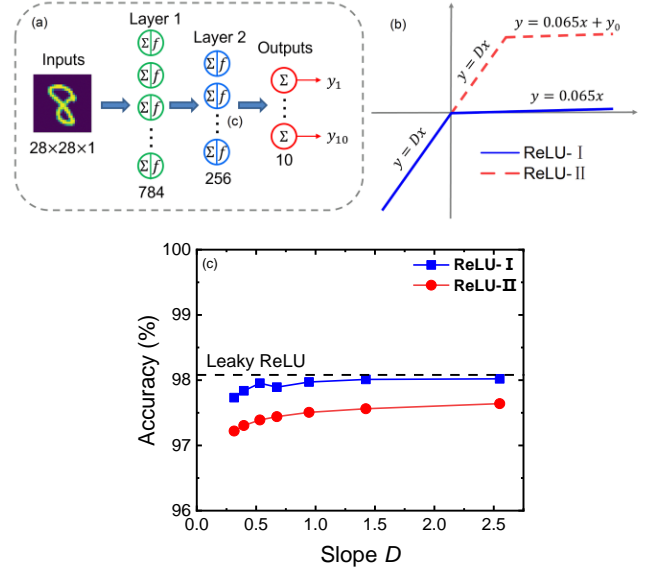

Fig. 4. (a) Deep neural network architecture. (b) Two types of optical ReLU-like functions. (c) Performance of the optical ReLU-like functions. The dash line indicates the accuracy of the standard leaky ReLU function.

The optical ReLU-like activation function is applied in a deep neural network with two fully-connected hidden layers. As shown in Fig. 4(a), the first layer consists of 784 nodes and the second one consists of 256 nodes. The performance of the neural network is tested in the classification task of the MNIST handwritten digits, where 60000 images in the database are used for training, and 10000 images are used for testing. We test two types of optical ReLU-like functions illustrated in Fig. 4(b): ReLU-I is defined as y=min($D$*x,0.065x), with $D$ being the measured slope in the range of [0.32, 2.55] (see Fig. 3(d)); ReLU-II is defined as y=min($D$*x, 0.065x+$y_0$), with x≥0. The constraint in ReLU-II is due to the fact that most optical neurons can only deal with positive input values. Figure 4(c) shows that the classification accuracies of both ReLU-I and ReLU-II generally rise with increasing slope $D$. The accuracy increases from 97.73% at $D$=0.32 to 98.02% at $D$=2.55 for ReLU-I, while from 97.22% to 97.64% for ReLU-II. As expectation, the accuracy of ReLU-I is slightly better than that of ReLU-II, because the input of ReLU-II is constrained to only positive values. In comparison, the classification of the standard leaky ReLU function (defined as y=max(0.01x, x)) is 98.08%. Therefore, the performances of both optical ReLU-like functions are comparable to that of the standard leaky ReLU function.

In conclusion, we have demonstrated an optical ReLU-like activation function based on a semiconductor laser with optical injection. The activation function occurs in a broad regime above the Hopf bifurcation and exists at different above-threshold pump currents. Therefore, the scheme is highly flexible for practical implementation. In particular, the output of the nonlinear neuron shares the same wavelength with the input, which is valuable for developing cascading optical neurons in deep neural networks. In addition, the slope of the ReLU-like function is reconfigurable through adjusting the detuning frequency. It is proved that the performance of the optical ReLU is comparable to that of the standard leaky ReLU. We believe that this work is helpful for developing all-optical neural networks with both optical synapses and optical neurons. Future work will investigate the response of the optical ReLU with modulated input signal. It is expected that the modulation speed can be comparable to the modulation bandwidth of the semiconductor laser.

**Funding.** Shanghai Science and Technology Program (21010502700), Natural Science Foundation of Shanghai (20ZR1436500).

**Acknowledgment.** The authors thank Prof. Sze-Chun Chan at City University of Hong Kong for very fruitful discussion.

**Disclosures.** The authors declare no conflicts of interest.

**Data availability.** All data generated or analyzed during this study are included in the published article.